\documentclass[12pt]{article}

\usepackage{arxiv}

\usepackage[utf8]{inputenc} 
\usepackage[T1]{fontenc}    
\usepackage{hyperref}       
\usepackage{url}            
\usepackage{booktabs}       
\usepackage{amsfonts}       
\usepackage{nicefrac}       
\usepackage{microtype}      
\usepackage{lipsum}
\usepackage{comment}
\usepackage{subcaption}
\usepackage{graphicx}
\usepackage{soul}           

\title{Neural networks based variationally enhanced sampling}

\author{
  Luigi Bonati\\
  Department of Physics, ETH Zurich, 8092 Zurich, Switzerland\\
  and Institute of Computational Sciences, Universit{\`a} della Svizzera italiana,\\via G. Buffi 13, 6900 Lugano, Switzerland \\
   \And
  Yue-Yu Zhang\\
  Department of Chemistry and Applied Biosciences, ETH Zurich, 8092 Zurich, Switzerland\\
  and Institute of Computational Sciences, Universit{\`a} della Svizzera italiana,\\via G. Buffi 13, 6900 Lugano, Switzerland \\
   \AND
  Michele Parrinello\\
  Department of Chemistry and Applied Biosciences, ETH Zurich, 8092 Zurich, Switzerland,\\
  Institute of Computational Sciences, Universit{\`a} della Svizzera italiana,\\via G. Buffi 13, 6900 Lugano, Switzerland, \\
  and Italian Institute of Technology, Via Morego 30, 16163 Genova, Italy
}

\defineheader{Bonati \textit{et al.}}

\begin{document}
\maketitle

\begin{abstract}
Sampling complex free energy surfaces is one of the main challenges of modern atomistic simulation methods. The presence of kinetic bottlenecks in such surfaces often renders a direct approach useless.  A popular strategy is to identify a small number of key collective variables and to introduce a bias potential that is able to favor their fluctuations in order to accelerate sampling. Here we propose to use machine learning techniques in conjunction with the recent variationally enhanced sampling method [\textit{Valsson and Parrinello, Phys. Rev. Lett. 2014}] in order to determine such potential. This is achieved by expressing the bias as a neural network. The parameters are determined in a variational learning scheme aimed at minimizing an appropriate functional. This required the development of a new and more efficient minimization technique. The expressivity of neural networks allows representing rapidly varying free energy surfaces, removes boundary effects artifacts and allows several collective variables to be handled.

\end{abstract}


Machine learning (ML) is changing the way in which modern science is conducted. Atomistic based computer simulations are no exceptions. Since the work of Behler and Parrinello \cite{Behler2007b} neural networks \cite{Behler2016,Zhang2017} or Gaussian processes \cite{Bartok2010b} are now almost routinely used to generate accurate potentials. More recently ML methods have been used to accelerate sampling, a crucial issue in molecular dynamics (MD) simulations, where standard methods allow only a very restricted range of time scales to be explored.
An important family of enhanced sampling methods is based on the identifications of suitable collective variables (CVs) that are connected to the slowest relaxation modes of the system \cite{Valsson2016}. Sampling is then enhanced by constructing an external bias potential $V(\mathbf{s})$  which depends on the chosen collective variables $\mathbf{s}$. In this context machine learning has been applied in order to identify appropriate CVs \cite{Chen2018,Schoberl2018,Hernandez2018,Wehmeyer2018a,Mendels2018} and to construct new methodologies \cite{Ribeiro2018,Mones2016,Galvelis2017,Sidky2018,Zhang2018c}. From these early experiences, it is also clear that ML applications can in turn profit from enhanced sampling \cite{Bonati2018}.

Here we shall focus on a relatively new method, called Variationally Enhanced Sampling (VES) \cite{Valsson2014}. In VES the bias is determined by minimizing a functional $\Omega=\Omega[V(\mathbf{s})]$. This functional is closely related to a Kullback Leibler divergence \cite{Invernizzi2017}. The bias that minimizes $\Omega$ is such that the probability distribution of $\mathbf{s}$ in the biased ensemble $P_V(\mathbf{s})$ is equal to a pre-assigned target distribution $p(\mathbf{s})$ . The method has been shown to be flexible \cite{Valsson2015,McCarty2015} and has great potential also for applications different from enhanced sampling. Examples of these heterodox applications are the estimation of the parameters of Ginzburg-Landau free energy models \cite{Invernizzi2017}, the calculation of critical indexes in second-order phase transitions \cite{Piaggi2016}, and the sampling of multithermal-multibaric ensembles \cite{Piaggi2019}.

Although different approaches have been suggested \cite{McCarty2016,Invernizzi2019}, the way in which VES is normally used is to expand $V(s)$ in a linear combination of orthonormal polynomials and use the expansion coefficients as variational parameters. In spite of its many successes, VES is not without problems. The choice of the basis set is often a matter of computational expediency and not grounded on physical motivations. Representing sharp features in the VES may require many terms in the basis set expansion. The number of variational parameters scales exponentially with the number of CVs and can become unmanageably large. Finally, non-optimal CVs may lead to very slow convergence.

In this paper, we use the expressivity \cite{Goodfellow-et-al-2016} of neural networks (NN) to represent the bias potential and a stochastic steepest descent framework for the determination of the NN parameters. In so doing we have developed a more efficient stochastic optimization scheme that can also be profitably applied to more conventional VES applications. 

\section*{Neural network based VES}
Before illustrating our method, we recall some ideas of CV-based enhanced sampling methods and particularly VES. 
\subsection*{Collective variables}
It is often possible to reduce the description of the system to a restricted number of collective variables $\mathbf{s}=\mathbf{s}(\mathbf{R})$, functions of the atomic coordinates $\mathbf{R}$, whose fluctuations are critical for the process of interest to occur. 
We consider the equilibrium probability distribution of these CVs as:
\begin{equation}
    P(\mathbf{s})=\int d\mathbf{R} \frac{e^{-\beta U(\mathbf{R})}}{Z} \delta \left(\mathbf{s}-\mathbf{s}(\mathbf{R})\right)
\end{equation}
where Z is the partition function of the system, $U(\mathrm{R})$ its potential energy and $\beta=(k_BT)^{-1}$ the inverse temperature. We can define the associated free energy surface (FES) as the logarithm of this distribution:
\begin{equation}
    F(\mathbf{s})=-\frac{1}{\beta} \log{P(\mathbf{s})}.
\end{equation}

Then an external bias is built as a function of the chosen CVs in order to enhance sampling. In umbrella sampling \cite{Torrie1977} the bias is static, while in Metadynamics \cite{Laio2002} is iteratively built as a sum of repulsive Gaussians centered on the points already sampled.

\begin{figure}[t!] 

\centering
\includegraphics[width=.4\linewidth]{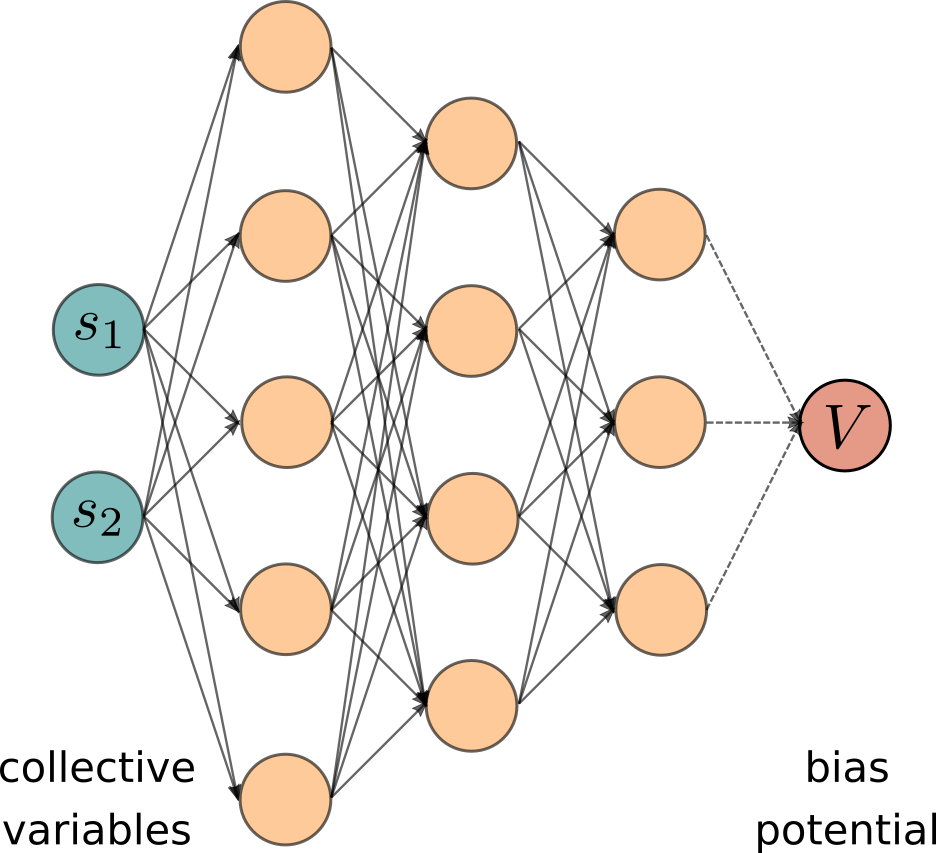} 
\caption{Neural network representation of the bias. The inputs are the chosen collective variables, whose values are propagated across the network in order to get the bias. The parameters are optimized according to the variational principle of eq. \ref{eq:omega}.}
\label{fig:nn}
\end{figure}

\subsection*{The variational principle}
In VES a functional of the bias potential is introduced: 
\begin{equation}
    \Omega[V]=\frac{1}{\beta} \log { \frac{\int {d\mathbf{s}e^{-\beta\left( F(\mathbf{s})+V(\mathbf{s})\right)}} }{\int {d\mathbf{s}e^{-\beta F(\bf{s})}}} } + \int {d\mathbf{s}\ p(\mathbf{s})V(\mathbf{s})}
    \label{eq:omega}
\end{equation}
where $p(\mathbf{s})$ is a chosen target probability distribution. The functional $\Omega$ is convex \cite{Valsson2014} and the bias that minimizes it is related to the free energy by the simple relation:
\begin{equation}
    F(\mathbf{s})= -V(\mathbf{s})-\frac{1}{\beta} \log{ p(\mathbf{s}) }
    \label{eq:fes}
\end{equation}
At the minimum the distribution of the CVs in the biased ensemble is equal to the target distribution:
\begin{equation}
    p_V(\mathbf{s})=p(\mathbf{s})
    \label{eq:p_eq}
\end{equation}
where $p_V(\mathbf{s})$ is defined as:
\begin{equation}
    p_V(\mathbf{s})=\frac{e^{-\beta \left( F(\mathbf{s}) + V(\mathbf{s})\right)}}{ \int d\mathbf{s} e^{-\beta ( F\left(\mathbf{s}) + V(\mathbf{s})\right)} }
\end{equation}
In other words, $p(\mathbf{s})$ is the distribution the CVs will follow when the $V(\mathbf{s})$ that minimizes $\Omega$ is taken as bias. This can be seen also from the perspective of the distance between the distribution in the biased ensemble and the target one. The functional can be indeed written as $\beta\Omega[V] = D_{KL}(p\ ||\ p_V) - D_{KL}(p\ ||\ P)$ \cite{Invernizzi2017}, where $D_{KL}$ denotes the Kullback-Leibler (KL) divergence.

\subsection*{The target distribution}
In VES an important role is played by the target distribution $p(\mathbf{s})$. A careful choice of $p(\mathbf{s})$ may focus sampling in relevant regions of the CVs space and in general accelerate convergence \cite{Shaffer2016a}. This freedom has been taken advantage of in the so called well-tempered VES \cite{Valsson2015}. In this variant, one takes inspiration from well-tempered metadynamics \cite{Barducci2008} and targets the distribution:
\begin{equation}
    p(\mathbf{s})=\frac{e^{-\beta F(\mathbf{s})/\gamma}}{\int d\mathbf{s}\ e^{-\beta F(\mathbf{s})/\gamma}} \propto \left[ P (\mathbf{s}) \right] ^{1/\gamma}
    \label{eq:target}
\end{equation}
where $P(\mathbf{s})$ is the distribution in the unbiased system and $\gamma>1$ is a parameter that regulates the amplitude of the $\mathbf{s}$ fluctuations. This choice of $p(\mathbf{s})$ has proven to be highly efficient \cite{Valsson2015}. Since at the beginning of the simulation $F(\mathbf{s})$ is not known, $p(\mathbf{s})$ is determined in a self-consistent way. Thus also the target evolves during the optimization procedure. In this work we update the target distribution every iteration, although less frequent updates are also possible

\subsection*{Neural network representation of the bias}
The standard practice of VES has been so far of expanding linearly $V(\mathbf{s})$ on a set of basis functions and use the expansion coefficients as variational parameters. Here this procedure is circumvented as the bias is expressed as a deep neural network as shown in figure \ref{fig:nn}. We call this variant DEEP-VES. The inputs of the network are the chosen CVs and this information is propagated to the next layers through a linear combination followed by the application of a non-linear function $a$ \cite{Goodfellow-et-al-2016}:
\begin{equation}
    \mathbf{x}^{l+1} = a (\mathbf{w}^{l+1} \mathbf{x}^{l}+\mathbf{b}^{l+1})
\end{equation}
Here the non linear activation function is taken to be a rectified linear unit. In the last layer only a linear combination is done, and the output of the network is the bias potential. 

We are employing NN since they are smooth interpolators. Indeed the NN representation ensures that the bias is continuous and differentiable. The external force acting on the i-th atom can be then recovered as: 
\begin{equation}
    \mathbf{F}_i=-\nabla_{\mathbf{R}_i} V = -\sum_{j=1} ^n \frac{\partial V}{\partial s_j} \nabla_{\mathbf{R}_i} s_j
\end{equation}
where the first term is efficiently computed via back-propagation. 
The coefficients $\{w^{i}\}$ and $\{b^{i}\}$ that we lump in a single vector $\mathbf{w}$ will be our variational coefficients. With this bias representation, the functional $\Omega[V]$ becomes a function of the parameters $\mathbf{w}$. Care must be taken to preserve the symmetry of the CVs, such as the periodicity. In order to accelerate convergence, we also standardize the input to have mean zero and variance one \cite{LeCun2012}. 

\begin{figure*}[ht!]
\begin{center}
\includegraphics[width=\textwidth]{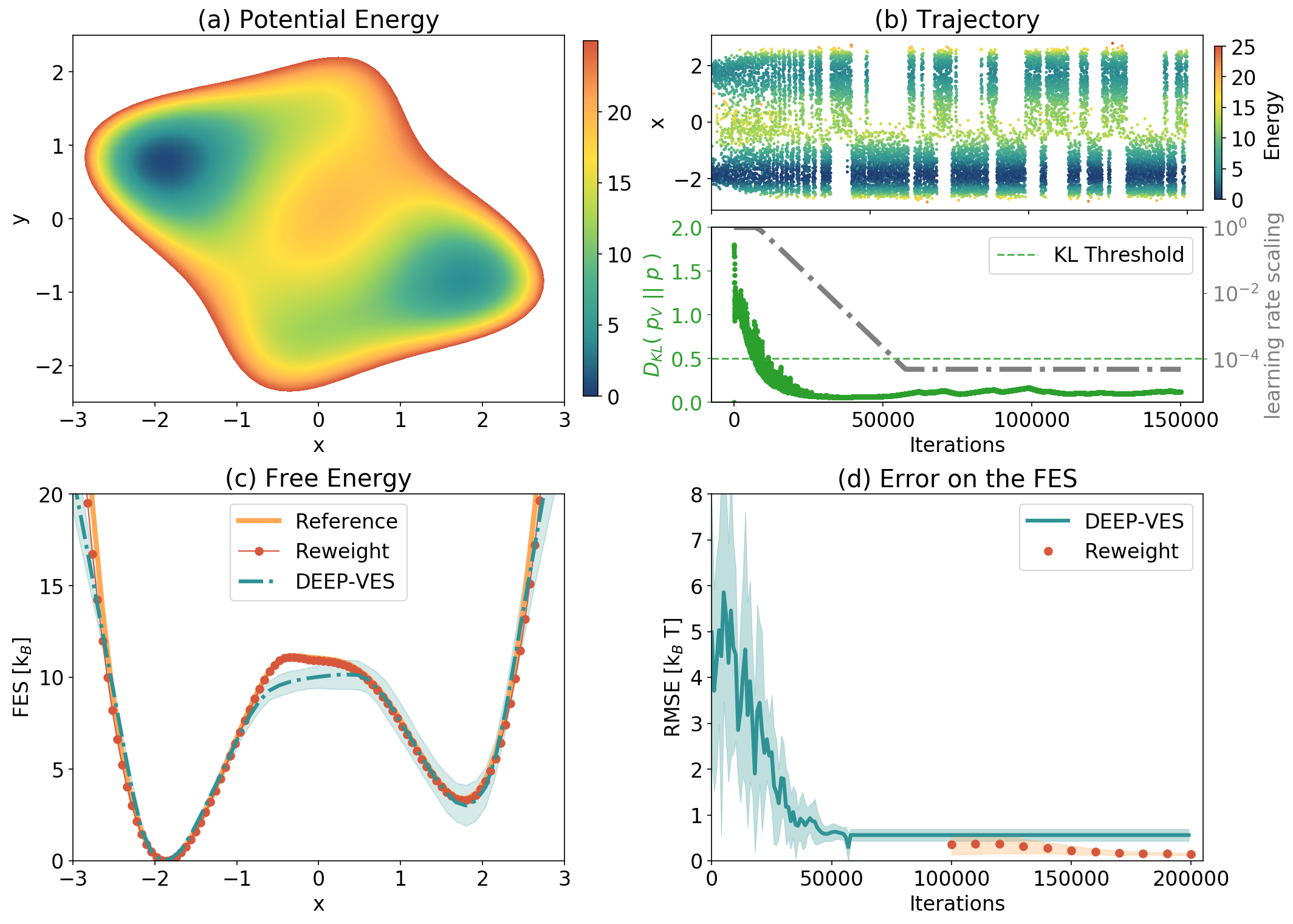}
\end{center}
\caption{Results for the 2D model. (a) Potential energy surface of the model. (b) Upper panel: CV evolution as a function of the number of iterations. The points are colored according to their energy value. Lower panel: evolution of the KL divergence between the bias and the target distribution (green) and learning rate scaling factor (grey). When the KL divergence is lowered below the threshold value the learning rate is decreased exponentially until it is practically zero. (c) Free energy profiles obtained from the neural network and with the reweighting procedure, compared with the reference obtained by integrating the model. For the DEEP-VES curve also an estimate of the error is given by averaging the results of 8 different simulations. The lack of accuracy on the left shoulder is a consequence of the combination of the sub-optimal character of the chosen variable and the strength of the bias, which creates new pathways other than the one of minimum energy. These artifacts are removed by performing the reweight with the static bias potential $V_s$. (d) Root mean square error (RMSE) on the FES computed in the regions of 10 $k_B T$ from the reference minimum.}
\label{fig:model}
\end{figure*}

\subsection*{The optimization scheme}
As in all neural network applications, training plays a crucial role. The functional $\Omega$ can be considered a scalar loss function with respect to the set of parameters $\mathbf{w}$. We shall evolve the parameters following the direction of the $\Omega$ derivatives:
\begin{equation}
    \frac{\partial \Omega}{\partial \mathbf{w}} = - \left\langle\frac{\partial V}{\partial \mathbf{w}} \right\rangle_{V} +\left\langle\frac{\partial V}{\partial \mathbf{w}} \right\rangle_{p}
    \label{eq:grad}
\end{equation}
where the first average is performed over the system biased by $V(\mathbf{s})$ and the second over the target distribution $p(\mathbf{s})$. It is worth noting that the form of the gradients is analogous to unsupervised schemes in energy-based models, where the network is optimized to sample from a desired distribution \cite{lecun2006tutorial}.

At every iteration, the sampled configurations are used to compute the first term of the gradient.
The second term, which involves an average over the target distribution, can be computed numerically when the number of CVs is small or using a Monte Carlo scheme when the number of CV is large. In doing so, the exponential growth of the computational cost with respect to the number of CV can be circumvented.
This scheme allows the neural network to be optimized on-the-fly, without the need to stop the simulation and fit the network to a given dataset.
In this respect, the approach used here resembles a reinforcement learning algorithm. The potential $V(\mathbf{s})$ can be regarded indeed as a stochastic policy that drives the system to sample the target distribution in the CVs space. This is a type of variational learning akin to what is done in applications of neural networks to quantum physics \cite{Carleo2017}.

Since the calculation the gradients requires performing statistical averages, a stochastic optimization method is necessary. In standard VES applications the method of choice has so far been the one of Bach and Moulines \cite{Bach2013}, that allows reaching with high accuracy the minimum, provided that the CVs are of good quality. Applying the same algorithm to NNs is rather complex. Therefore we espouse the prevailing attitude in the neural networks community. Namely, we do not aim at reaching the minimum rather at determining a value of the bias $V_s (s)$ that is close enough to the optimum value to be useful. In the present context this means that a run biased by $V_s (s)$ can be used to compute the Boltzmann equilibrium averages from the well-known umbrella sampling-like formula:

\begin{equation}
    \left\langle O(\mathbf{R}) \right\rangle = \frac{\left\langle O(\mathbf{R}) e^{\beta V_s(s(\mathbf{R}))}\right\rangle_{V_{s}}}{\left\langle e^{\beta V_s(s(\mathbf{R}))} \right\rangle_{V_s}}.
    \label{eq:rew}
\end{equation}

In order to measure the progression of the minimization towards eq. \ref{eq:p_eq} we monitor at iteration step $n$ an approximate KL divergence between the running estimates of the biased probability $p_V ^{(n)} (s)$ and the target $p ^{(n)} (s)$:
\begin{equation}
    D_{KL} ^{(n)} (\ p_V\ ||\ p\ ) = \sum_{\mathbf{s}} p_V ^{(n)} (\mathbf{s}) \log { \frac{p_V ^{(n)} (\mathbf{s})}{p^{(t)}(\mathbf{s})} } 
    \label{eq:kl}
\end{equation}
These quantities are estimated as exponentially decaying averages. In such a way only a limited number of configurations contribute to the running KL divergence.

The simulation is thus divided into three parts. In the first one the ADAM optimizer \cite{Kingma2014} is used, until a pre-assigned threshold value $\epsilon$ for the distance $D_{KL}(\ p_V\ ||\ p\ )$ is reached. From then on the learning rate is exponentially brought to zero provided that $D_{KL}$ remains below $\epsilon$. From this point on, the network is no longer updated and statistics is accumulated using eq. \ref{eq:rew}. The longer the second part, the better the bias potential $V_s$, and the shorter phase three need to be. Faster decay times instead need to be followed by longer statistics accumulation runs. However, in judging the relative merits of these strategies it must be taken into account that the third phase in which the bias is kept constant involves a minor number of operations and it is therefore much faster.

\begin{figure*}[t!]
\begin{center}
\includegraphics[width=\linewidth]{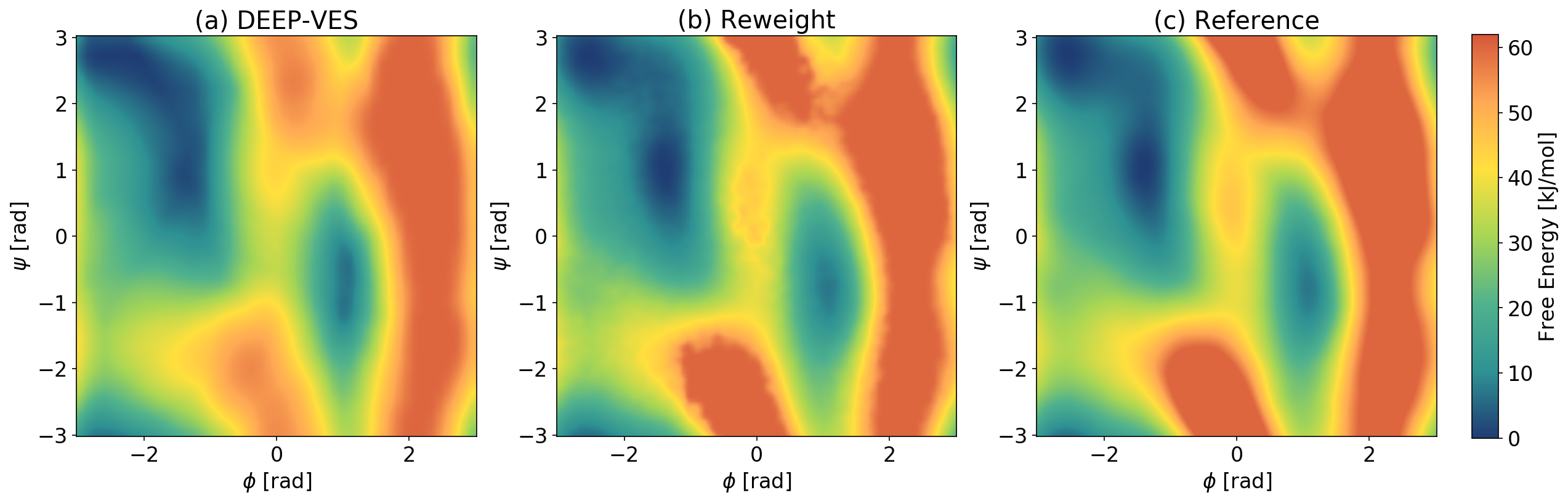}
\end{center}
\caption{Alanine Dipeptide free energy results. (a) DEEP-VES representation of the bias. (b) FES profile obtained with reweighting. (c) Reference from a 100 ns meta dynamics simulation. The main features of the FES are captured by the the neural network, which allow for an efficient enhanced sampling of the conformations space. Finer details can be easily recovered with the reweighting procedure. }
\label{fig:ala2-fes}
\end{figure*}

\section*{Results}

\subsection*{Wolfe-Quapp potential}
We first focus on a toy-model, namely the two-dimensional Wolfe-Quapp potential, rotated as in ref. \cite{Invernizzi2019}. This is shown in figure \ref{fig:model}a. The reason behind this choice is that the dominant fluctuations that lead from one state to the other are in an oblique direction with respect to $x$ and $y$. We choose on purpose to use only $x$ as a CV in order to exemplify the case of a sub-optimal CV. This is representative of what happens in the practice when, more often than not, some slow degrees of freedom are not fully accounted for \cite{Invernizzi2019}. 

We use a three-layer neural network with [48,24,12] nodes, resulting in 1585 variational parameters, which are updated every 500 steps. The KL divergence between the biased and the target distribution is computed with an exponentially decaying average with a time constant of $5\cdot10^4$ iterations. 
The threshold $\epsilon$ is set to 0.5 and the learning rate decay time to $5\cdot10^3$ iterations. The results are robust with respect to the choice of these parameters, see the Supporting Information (SI).

In the upper right panel (figure \ref{fig:model}b) we show the evolution of the CV as a function of the number of iterations. 
At the beginning, the bias changes very rapidly with large fluctuations that help to explore the configuration space. It should be noted that the combination of a sub-optimal CV and a rapidly varying potential might lead the system to pathways different from the lowest energy one. For this reason, it is important to slow down the optimization in the second phase, where the learning rate is exponentially lowered until it is practically zero. When this limit is reached the frequency of transitions becomes lower, reflecting the sub-optimal character of the CV \cite{Invernizzi2019}, as can be seen in fig. \ref{fig:model}b.

Although the bias itself is not yet fully converged, still the main features of the FES are captured by $V_s(\mathbf{s})$ (fig. \ref{fig:model}c). This is ensured by the fact that while decreasing the learning rate the running estimate of the KL divergence must stay below the threshold $\epsilon$, otherwise the optimization proceeds with a constant learning rate. This means that the final potential is able to promote efficiently transitions between the metastable states. The successive static reweighting refines the FES and leads to a very accurate result (fig. \ref{fig:model}d), removing also the artifacts caused in the first part by the rapidly varying potential.

\subsection*{Alanine dipeptide and tetrapeptide}
As a second example we consider the case of two small peptides, Alanine Dipeptide and Alanine Tetrapeptide in vacuum, which are often used as a benchmark for enhanced sampling methods. We will refer to them as Ala2 and Ala4 respectively. 
Their conformations can be described in terms of the Ramachandran angles $\phi_i$ and $\psi_i$, where the first ones are connected to the slowest kinetic processes. The smaller Ala2 has only one pair of such dihedral angles, which we will denote as $\{\phi,\psi\}$, while Ala4 has three pairs of backbone angles denoted by $\{ \phi_i,\psi_i\}$ with $i=1,2,3$. 

\begin{figure}[h!]
\centering
  \begin{subfigure}[h!]{0.45\textwidth}
    \includegraphics[width=0.85\textwidth]{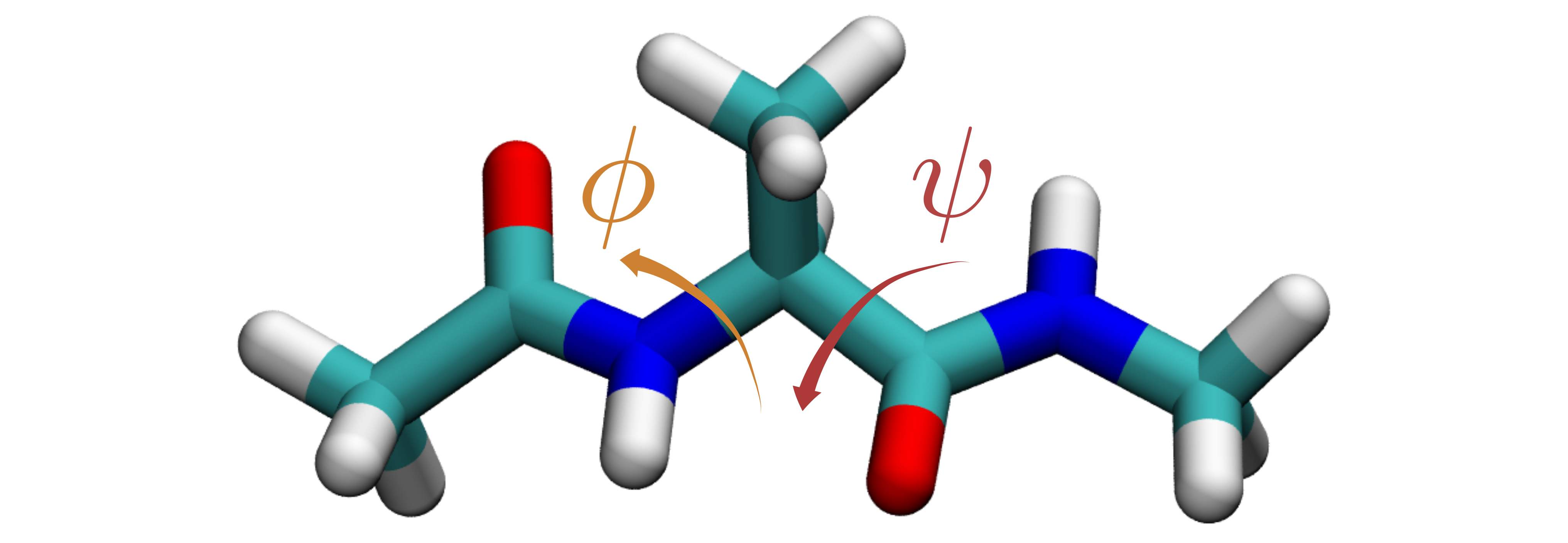}
    \caption{Alanine dipeptide}
    \label{fig:1}
  \end{subfigure}
  \begin{subfigure}[h!]{0.45\textwidth}
    \includegraphics[width=\textwidth]{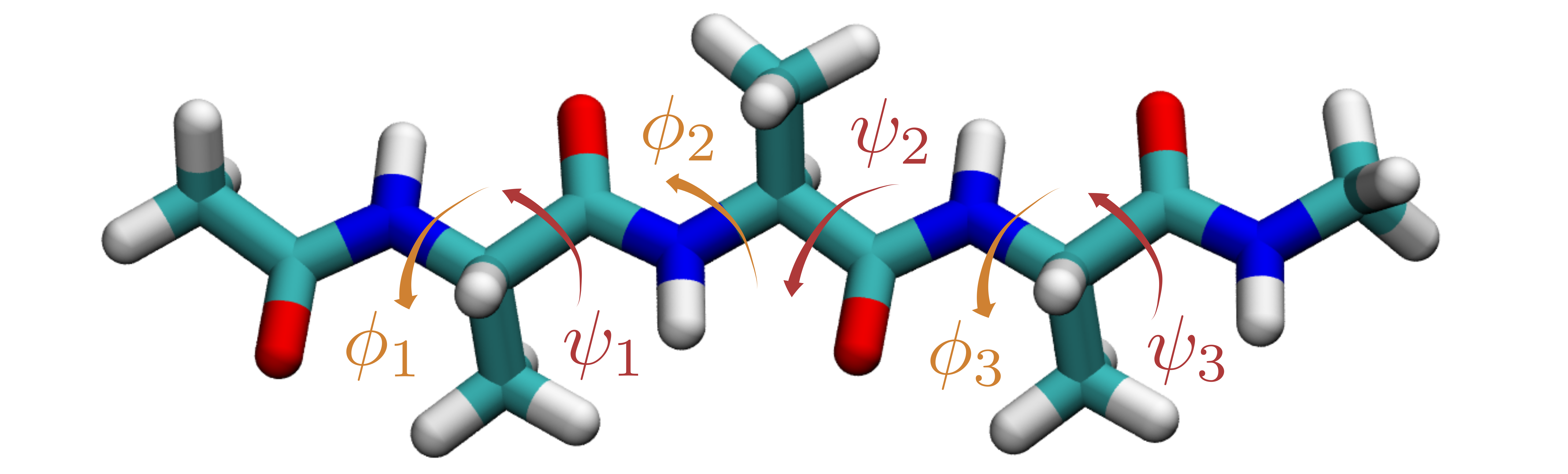}
    \caption{Alanine tetrapeptide}
    \label{fig:2}
  \end{subfigure}
\end{figure}

We want to show here the usefulness of the flexibility provided by DEEP-VES in different systems. For this purpose, we use the same architecture and optimization scheme as in the previous example. We only decrease the decay time of the learning rate in the second phase since the diedral angles $\{\phi,\psi\}$ are known to be a good set of CVs.
In order to enforce the periodicity of bias potential, the angles are first transformed in their sines and cosines: $\{\phi,\psi\}\rightarrow\{\cos(\phi),\sin(\phi),\cos(\psi),\sin(\psi)\}$.

The simulation of Ala2 reached the KL divergence threshold after 3 ns, and from there on the learning rate was exponentially decreased. The NN bias was no longer updated after 12 ns. At variance with the first example, when the learning is slowed down and even when it is stopped the transition rate is not affected: this is a signal that our set of CVs is good (see SI). In figure \ref{fig:ala2-fes} we show the free energy profiles obtained from the NN bias following eq. \ref{eq:fes} and the one recovered with the reweighting procedure of eq. \ref{eq:rew}. We compute the root mean square error (RMSE) on the FES up to 20kJ/mol as in ref. \cite{Valsson2015}. The errors are 1.5 and 0.45 kJ/mol, corresponding to 0.6 and 0.2 $k_B T$, both well below the threshold of chemical accuracy.

\begin{figure*}[b!]
\begin{center}
\includegraphics[width=\linewidth]{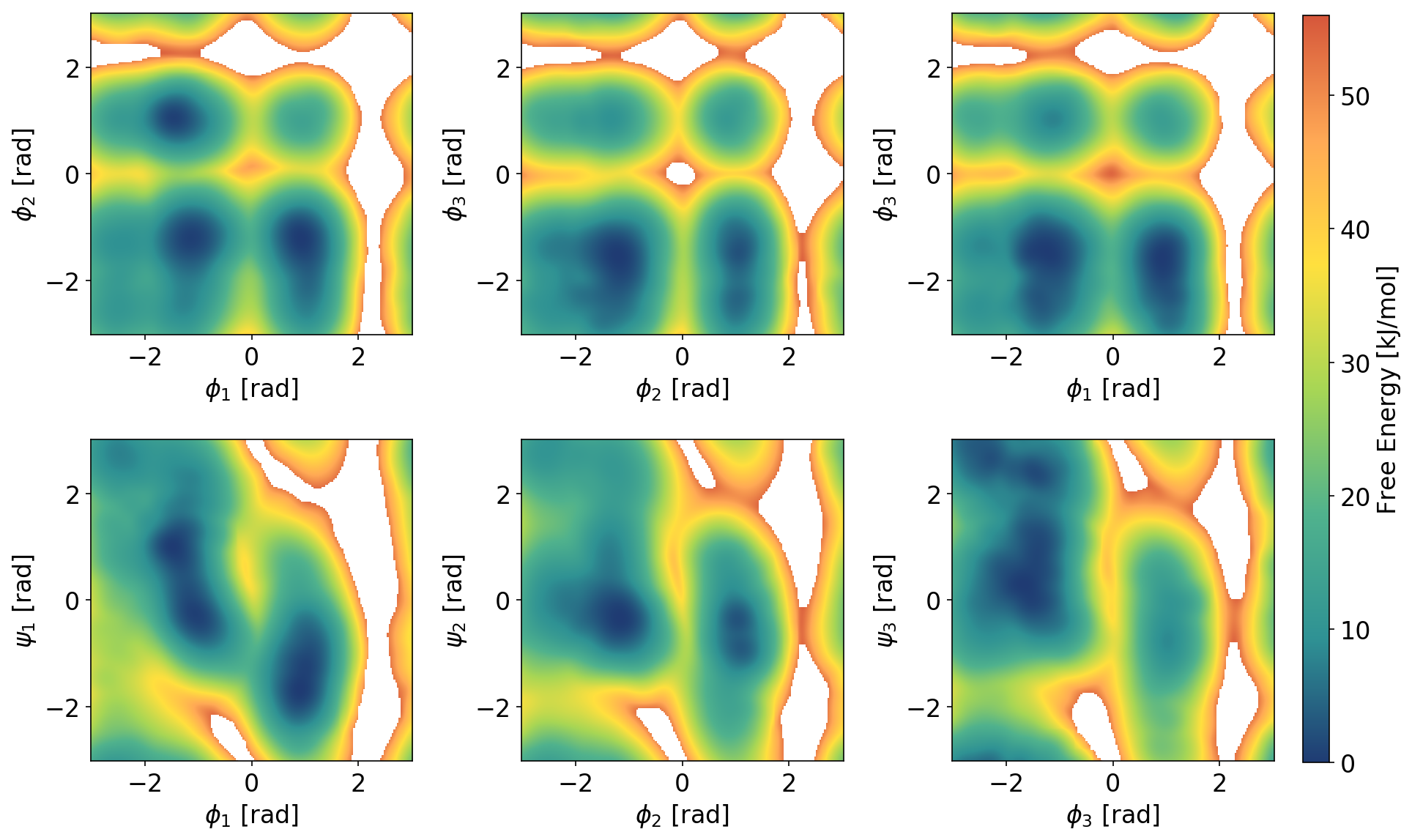}
\end{center}
\caption{Two-dimensional free energy surfaces of Ala4 obtained with the reweighting, as a function of the three dihedral angles $\{\phi_1,\phi_2,\phi_3\}$. The FES in terms of other pairs as well the comparison with the references is reported in the SI.}
\label{fig:ala4-fes}
\end{figure*}

In order to exemplify the ability of DEEP-VES to represent functions of several variables we study the FES of Ala4 in terms of its six dihedral angles $\{\phi_1,\psi_1,\phi_2,\psi_2,\phi_3,\psi_3\}$ (fig. \ref{fig:ala4-fes}). In the case of a small number of variables the second integral on the right-hand side of eq. \ref{eq:grad} is calculated numerically on a grid. This is clearly not possible in a six-dimensional space, thus we resort to a Monte Carlo estimation based on the Metropolis algorithm \cite{metropolis1953equation}. Convergence is helped by the fact that the target $p(\mathbf{s})$ is broadened by the well tempered factor $\gamma$ (eq. \ref{eq:target}). In this case, we cannot use eq. \ref{eq:kl} to monitor convergence. However, as the number of CV treated increases the probability of the space spanned by the CVs of covering all the relevant slow modes is very high. Thus less attention needs to be devoted to the optimization schedule. The transition from the constant learning rate phase to the static bias region can be initiated after multiple recrossings between metastable states are observed. In order to verify the accuracy of these results, we compared the free energy profiles obtained from the reweight with the results from metadynamics and its Parallel Bias version \cite{Pfaendtner2015} (SI).

\subsection*{Silicon crystallization}

In the last example, the phase transition of silicon from liquid to solid at ambient pressure is studied. This is a complex phenomenon, characterized by a high barrier around 80 $k_B T$, and several CVs have been proposed to enhance this process. The crystallization and the melting process are marked by a local character. As a consequence, many CVs are defined in terms of per-atom crystallinity measures. Then the number of solid-like atoms is used as collective variable. In terms of these variables, the free energy profile is characterized by very narrow minima, around 0 and $N_{atoms}$. In order to enhance in an efficient way the fluctuations in such space one often needs to employ variants of metadynamics \cite{Branduardi2012} or a high order basis set expansion of the VES.

\begin{figure}[h!]
\begin{center}
\includegraphics[width=0.65\linewidth]{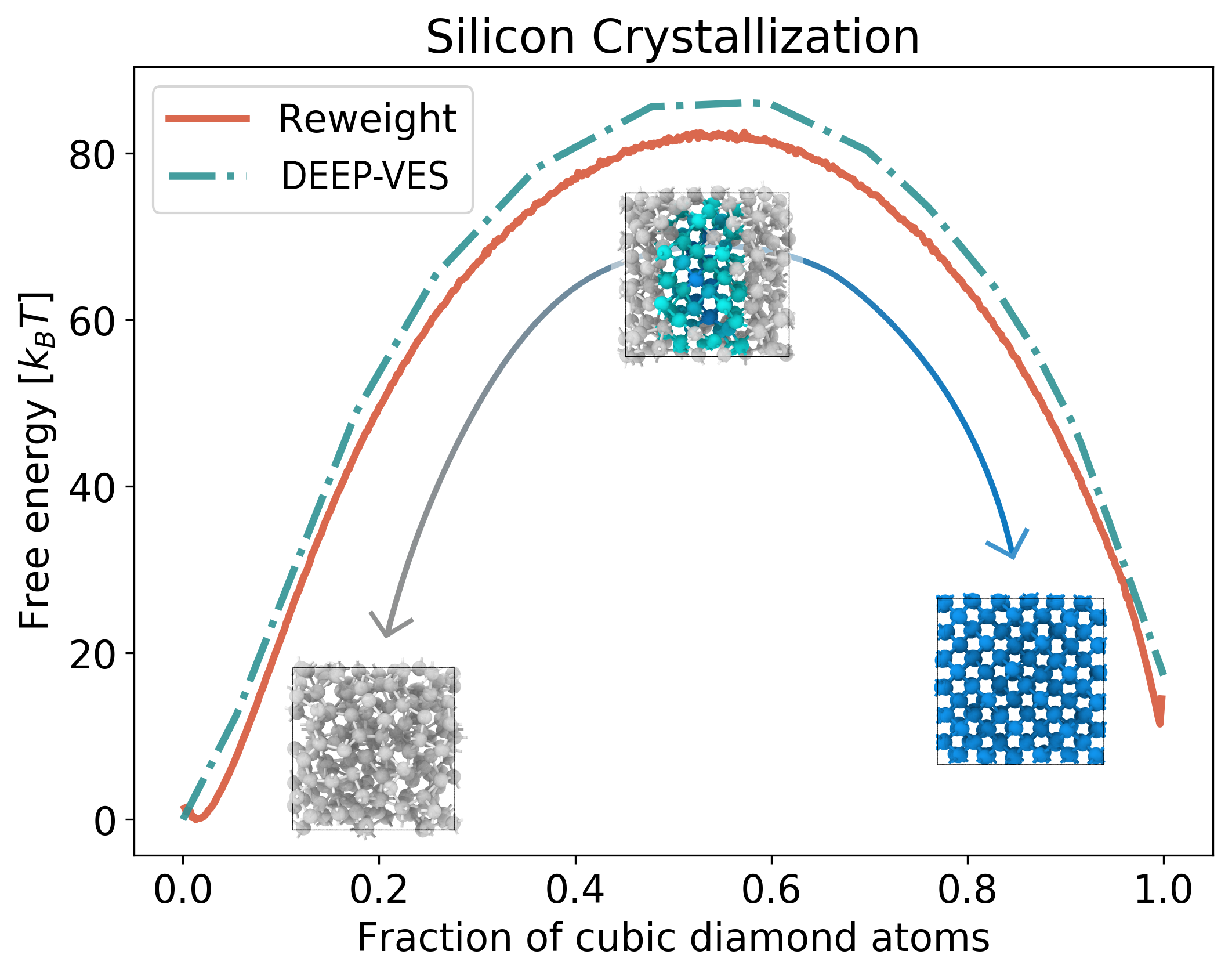}
\end{center}
\caption{Free energy surface of silicon crystallization, in terms of the number of cubic diamonds atoms in the system. Snapshots of the two minima, as well as the transition state are also shown. The reweight is obtained using $V_s$. Regions of the FES greater than 50 $kJ/mol$ are not shown.}
\label{fig:si-fes}
\end{figure}

This can be easily addressed using the flexibility of neural networks, which we use here to represent rapidly varying features of the FES and to avoid boundaries artifacts. In figure \ref{fig:si-fes} we show the free energy profile learnt by the NN as well the one recovered by reweighting. Even in the presence of very sharp boundaries, the NN bias is able to learn a good $V_s$, which allows for going back and forth efficiently between the two states and for recovering a good estimate of the FES in the subsequent reweight. The KL divergence threshold is reached in 10 ns, and the bias is no longer updated after 30ns. 

\subsection*{Choice of the parameters}
After presenting the results, we would like to spend a few words on the choice of parameters. This procedure requires setting three parameters, namely the time scale for calculating the KL divergence, the KL threshold and the decay time for the learning rate. Our experience is that if the chosen set of CVs is good this choice has a limited impact on the simulation result. However, in the case of non-optimal CVs, more attention is needed. 

The KL divergence should be calculated on a time scale in which the system samples the target distribution, so the greater the relaxation time of the neglected variables, the greater this scale should be. However, this is only used to monitor convergence and therefore it is possible to choose a higher value without compromising speed. A similar argument applies to the decay constant of the learning rate, but in the spirit of reaching a constant potential quickly, one tries to keep it lower (in the order of thousands of iterations). Finally, we found that the protocol is robust with respect to the epsilon parameter of the KL divergence threshold, provided that it is chosen in a range of values around 0.5. In the case of FES with larger dimensionality, it may be appropriate to increase this parameter in order to reach rapidly a good estimate of $V_s$. In the supplementary information, we report a study of the influence of these parameters in the accuracy of the bias learnt and the time needed to converge for the Wolfe-Quapp potential.

\section*{Conclusions}
In this work, we have shown how the flexibility of neural networks can be used to represent the bias potential and the FES as well, in the context of the VES method. Using the same architecture and similar optimization parameters we were able to deal with different physical systems and FES dimensionalities.  This includes also the case in which some important degree of freedom is left out from the set of enhanced CVs. 
The case of Alanine Tetrapeptide with a 6-dimensional bias already shows the capability of DEEP-VES of dealing with high-dimensional FES.
We plan to extend this to even higher dimensional landscapes, where the power of NNs can be fully exploited.

Our work is an example of a variational learning scheme, where the neural network is optimized following a variational principle. Also, the target distribution allows an efficient sampling of the relevant configurational space, which is particularly important in the optics of sampling high dimensional FES. 

In the process, we have developed a minimization procedure alternative to that of ref. \cite{Valsson2014}, which globally tempers the bias potential based on a Kullback Leibler divergence between the current and the target probability distribution. We think that also conventional VES can benefit from our new approach and that we have made another step in the process of sampling complex free energy landscapes. Future goals might include developing a more efficient scheme to exploit the variational principle in the context of NNs, as well as learning not only the bias but also the collective variables on-the-fly. This work allows also tapping into the immense literature on machine learning and neural networks for the purpose of improving enhanced sampling.

\clearpage
\section*{Materials and methods}
The VES-NN is implemented on a modified version of PLUMED2 \cite{Tribello2014}, linked against LibTorch (PyTorch C++ library). We plan to release the code in the open-source PLUMED package in the future. We take care of the construction and the optimization of the neural network with LibTorch. The gradients of the functional with respect to the parameters are computed inside PLUMED according to eq. \ref{eq:grad}. The first expectation value is computed by sampling, while the second is obtained by numerical integration over a grid or with Monte Carlo techniques. In the following we report the setup for the VES-NN and the simulations for all the examples reported in the paper.

\subsection*{Wolfe-Quapp potential}
The Wolfe-Quapp potential is a fourth-order polynomial: $U(x,y)=x^4 + y^4-2\ x^2-4\ y^2 +xy +0.3\ x + 0.1 y$. We rotated it by an angle $\theta=-3/20\ \pi$ in order to change the direction of the path connecting the two minima. A langevin dynamics is run with PLUMED using a timestep of 0.005, a target temperature of 1 and a friction parameter equal to 10 (in terms of natural units). The biasfactor for the well-tempered distribution is equal to 10. A neural network composed by [48,24,12] nodes is used to represent the bias. The learning rate is equal to 0.001. An optimization step of the NN is performed every 500 timesteps. The running KL divergence is computed on a timescale of $5\cdot10^4$ iterations. The threshold is set to $\epsilon=0.5$ and the decay constant for the learning rate is set to $5\cdot10^3$ iterations. The grid for computing the target distribution integrals has 100 bins. The parameters of the NN are the kept the same also for the following examples, unless otherwise stated. 

\subsection*{Peptides}
For the Alanine dipeptide (Ace-Ala-Nme) and tetrapeptide (Ace-Ala3-Nme) simulations we use GROMACS  \cite{VanDerSpoel2005} patched with PLUMED. The peptides are simulated in the NVT ensemble using the Amber99-SB force field \cite{Hornak2006} with a time step of 2 fs. The target temperature of 300K is controlled with the velocity rescaling thermostat \cite{Bussi07}. For Ala2 we use the following parameters: decay constant equal to $10^3$ and threshold equal to $\epsilon=0.5$. A grid of 50x50 bins is used. For Ala4 the Metropolis algorithm is used to generate a set of points according to the target distribution. At every iteration 25000 points are sampled. The learning rate is kept constant for the first 10 ns, and then decreased with a decay time of $2\cdot10^3$ iterations. In both cases the biasfactor is equal to 10.

\subsection*{Silicon}
For the silicon simulations we use LAMMPS \cite{Plimpton2007a} patched with PLUMED, employing the Stillinger and Weber potential \cite{Stillinger1985}. A 3x3x3 supercell (216 atoms) is simulated in the NPT ensemble with a timestep of 2 fs. The temperature of the thermostat \cite{Bussi07} is set to 1700K with a relaxation time of 100 fs, while the values for the barostat \cite{Martyna1994a} are 1 atm and 1 ps. The CV used is the number of cubic diamond atoms in the system, defined according to ref. \cite{Piaggi2019b}.The decay time for the learning rate is $2\cdot10^3$, and a grid of 216 bins is used. The biasfactor used is equal to 100.

\clearpage

\bibliographystyle{unsrt}  
\bibliography{main}  

\end{document}